\documentclass{sig-alternate}

\usepackage{xcolor}

\usepackage{hyperref}
\usepackage[noend]{algpseudocode}
\usepackage{algorithm}
\usepackage{algorithmicx}

\usepackage{microtype}
\usepackage{times,epsf,epsfig,textcomp,amsmath,amssymb}
\usepackage{wrapfig,moreverb}
\usepackage{comment}
\usepackage{array}
\usepackage{colortbl}
\usepackage{url}
\usepackage{color, soul}

\usepackage{graphics}

\usepackage[font={small,bf}]{caption}
\usepackage{subcaption}

\usepackage{multirow}
\usepackage{capt-of}
\usepackage{enumitem}
\usepackage{siunitx}
\usepackage{listings}
\usepackage{stmaryrd}

\usepackage{bm}
\usepackage[mathscr]{euscript}
\usepackage{xspace}

\usepackage{helvet}
\usepackage{tgheros}

\newcolumntype{L}[1]{>{\raggedright\let\newline\\\arraybackslash\hspace{0pt}}m{#1}}
\newcolumntype{C}[1]{>{\centering\let\newline\\\arraybackslash\hspace{0pt}}m{#1}}
\newcolumntype{R}[1]{>{\raggedleft\let\newline\\\arraybackslash\hspace{0pt}}m{#1}}

\usepackage{amsthm}

\theoremstyle{plain}

\theoremstyle{definition}
\definecolor{lightgray}{gray}{0.6}
\definecolor{lightblue}{rgb}{0.9,0.9,1}

\newcommand{\remove}[1]{}

\newcommand{\dstate}[1]{\texttt{\small #1}}
\newcommand{\ustate}[1]{\textsf{\small #1}}

\let\oldhat\hat
\renewcommand{\hat}[1]{\oldhat{\mathbf{#1}}}

\newcommand{\name}{POD\xspace}

\begin{document}

% Copyright
% \setcopyright{acmcopyright}
%\setcopyright{acmlicensed}
%\setcopyright{rightsretained}
%\setcopyright{usgov}
%\setcopyright{usgovmixed}
%\setcopyright{cagov}
%\setcopyright{cagovmixed}

% DOI
% \doi{10.475/123_4}

% ISBN
% \isbn{123-4567-24-567/08/06}

%Conference
% \conferenceinfo{PLDI '13}{June 16--19, 2013, Seattle, WA, USA}

% \acmPrice{\$15.00}

%
% --- Author Metadata here ---
% \conferenceinfo{WOODSTOCK}{'97 El Paso, Texas USA}
%\CopyrightYear{2007} % Allows default copyright year (20XX) to be over-ridden - IF NEED BE.
%\crdata{0-12345-67-8/90/01}  % Allows default copyright data (0-89791-88-6/97/05) to be over-ridden - IF NEED BE.
% --- End of Author Metadata ---

%different conventions exist about capitalizing "that," but usually it's capitalized
%the first word after colons in titles should ALWAYS be capitalized
\title{POD: A Smartphone That Flies\vspace{-0.5em}}
% \subtitle{[Extended Abstract]


\numberofauthors{3} %  in this sample file, there are a *total*

% % You can go ahead and credit any number of authors here,
% % e.g. one 'row of three' or two rows (consisting of one row of three
% % and a second row of one, two or three).
% %
% % The command \alignauthor (no curly braces needed) should
% % precede each author name, affiliation/snail-mail address and
% % e-mail address. Additionally, tag each line of
% % affiliation/address with \affaddr, and tag the
% % e-mail address with \email.
% %
% % 1st. author

% \author[1]{Guojun Chen}
% \author[1]{Noah Weiner}
% \author[1]{Lin Zhong}

% \affil{Yale University}
\author{
\alignauthor Guojun Chen\\
       \affaddr{Yale University}\\
       \email{guojun.chen@yale.edu}
% 2nd. author
\alignauthor Noah Weiner\\
       \affaddr{Yale University}\\
       \email{noah.weiner@yale.edu}
% 3rd. author
\alignauthor Lin Zhong\\
       \affaddr{Yale University}\\
       \email{lin.zhong@yale.edu}\\
}
% There's nothing stopping you putting the seventh, eighth, etc.
% author on the opening page (as the 'third row') but we ask,
% for aesthetic reasons that you place these 'additional authors'
% in the \additional authors block, viz.

% Just remember to make sure that the TOTAL number of authors
% is the number that will appear on the first page PLUS the
% number that will appear in the \additionalauthors section.

\maketitle




%
%  Use this command to print the description
%
% \printccsdesc

% We no longer use \terms command

% \keywords{ACM proceedings; \LaTeX; text tagging}

%% Abstract
%% Note: \begin{abstract}...\end{abstract} environment must come before \maketitle command\

% !TEX root = main.tex
\begin{abstract}
We present \name, a smartphone that flies, as a new way to achieve hands-free, eyes-up mobile computing. Unlike existing drone-carried user interfaces, \name features a smartphone-sized display and the computing and sensing power of a modern smartphone. 
We share our experience in prototyping \name, discuss the technical challenges facing it, and describe early results toward addressing them.
\end{abstract}

%% 2012 ACM Computing Classification System (CSS) concepts
%% Generate at 'http://dl.acm.org/ccs/ccs.cfm'.
% <concept_id>10011007.10011006.10011008</concept_id>
% <concept_desc>Software and its engineering~General programming languages</concept_desc>
% <concept_id>10003456.10003457.10003521.10003525</concept_id>
% <concept_desc>Social and professional topics~History of programming languages</concept_desc>
% <concept_significance>300</concept_significance>

% \ccsdesc[500]{Software and its engineering~General programming languages}
% \ccsdesc[300]{Social and professional topics~History of programming languages}
%% End of generated code

%% Keywords, a comma separated list
% \keywords{keyword1, keyword2, keyword3}  %% \keywords are mandatory in final camera-ready submission

%% \maketitle
%% Note: \maketitle command must come after title commands, author
%% commands, abstract environment, Computing Classification System
%% environment and commands, and keywords command.

\thispagestyle{empty}
\sloppy % Kevin: is there a better way to fix column wrapping?
%\setlength{\skip\footins}{5mm}

%\fontsize{10}{11.8}
%\selectfont

%NOTE: the abstract is above, before the \maketitle command
% !TEX root = ./main.tex 
\section{Introduction}\label{sec:intro}
Our driving vision is \emph{hands-free, eyes-up} mobile computing. That is, mobile users are able to interact with a computer without holding a device in hand and looking down at it.
The key to this vision is a user interface (both input and output) that does not require a human hand.
Wearable devices, such as Google Glass, are perhaps the most studied implementation of this type of mobile, hands-free user interface.

%%%% Explain our concept: DCUI
This paper explores an alternative idea for these new user interfaces: one carried by a small drone. Compared to wearable user interfaces, drone-carried UIs do not impose gadgets on the user's head or eyes. More importantly, they can go beyond the physical reach of the human body, empowering interesting use cases.

While there is a growing literature about human-drone interaction~\cite{funk2018hdi,tezza2019hdi,cauchard2021hdi}, this work envisions a drone-carried UI as rich as that of modern smartphones, called \name. Unlike previously studied drone-carried UIs, \name features a small (potentially foldable) display and allows bidirectional interactions with mobile users. 
We imagine that \name would rest on the user's shoulder most of the time and fly out occasionally to interact with the user only on demand. 
% Obviously, this is inspired in part by characters from popular culture, e.g., Jack the Monkey, a pet of Captain Barbossa, EVE from \emph{WALL-E}, as well as the Decepticon spies of Soundwave from \emph{Transformers}.

% \begin{figure}
%     \centering
%     \includegraphics[width=0.6\linewidth]{image/concept1.jpg}
%     \caption{(a) a \name rests on the user shoulder, being charged. (b) the \name flies out and presents it to the user when she answers a video call. As she moves around, the \name follows her.}
%     \label{fig:dcui}
% \end{figure}

%%% Summarize the challenges
Compared to wearable glasses and other drone-carried UIs, \name faces its own unique challenges that have not been addressed by the human-drone interaction community.
First, like all flying objects, drones are not steady.
This challenges the usability of an onboard display, especially if the drone needs to maintain a safe distance of one to two feet from the user's eyes.
Second, drones are noisy. Even the quietest drone operating at one or two feet away from the user produces noise at 70 dBA, which is equivalent to being on a busy street or near a vacuum cleaner~\cite{NoiseCDC}. 
Finally, because we intend \name to serve the user on the go, it must decide between following the user or hovering in place when the user moves around. This challenge represents a very specialized case of human-drone interaction and requires new investigation. 

\begin{figure}
    \centering
    \includegraphics[width=0.5\linewidth]{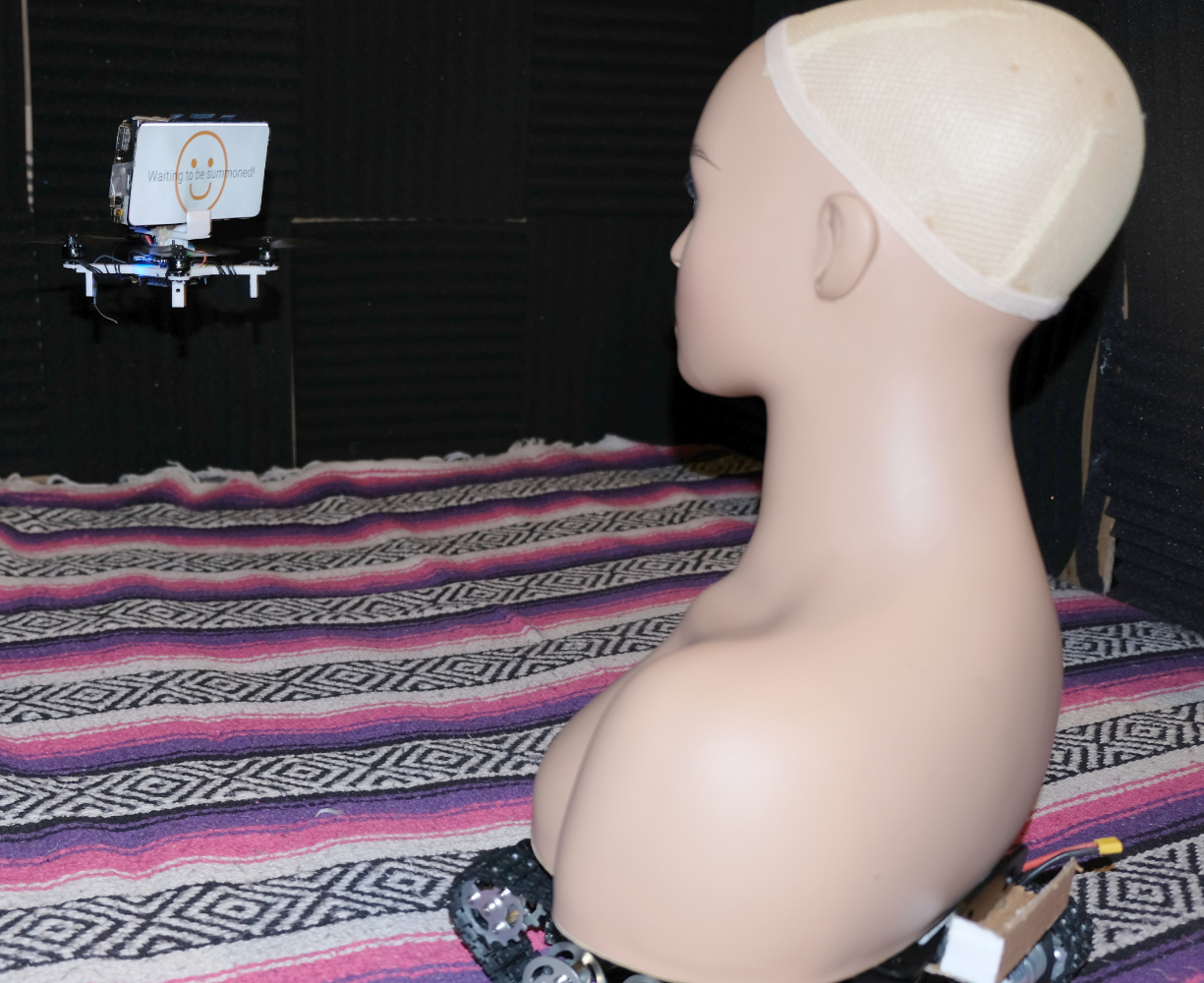}
    \caption{\name (top left corner) interacting with our test mannequin mounted on a Wi-Fi-controlled robotic chassis.}
    \label{fig:podamable}
    \vspace{-1ex}
\end{figure}

%%%% Describe our prototype: POD.
To address these challenges and experiment with \name, we have implemented a prototype. As shown in \autoref{fig:podamable}, it is based on a mechanical, electrical and computational partnership between a micro-controller-based drone and a stripped-down smartphone (Google Pixel 4). It has a flight time of four minutes, uses computer vision to track and follow its user, and harnesses the phone's sensors to improve the usability of its display.

This paper describes the \name prototype and shares our experience in building it (\S\ref{sec:system}). It presents our early results (\S\ref{sec:evaluation}) in addressing the challenges facing \name (\S\ref{sec:challenge}) and discusses interesting, novel use cases for it (\S\ref{sec:use}).

% \begin{figure}
%     \centering
%     \includegraphics[width=0.4\linewidth]{image/dcui.pdf}
%     \caption{\name is based on an in-house designed nano-drone and a Pixel 4.}
%     \label{fig:pod}
% \end{figure}
\section{The case for \name}
\label{sec:use}

\begin{figure*}[!ht]
\centering
    \begin{minipage}{0.45\textwidth}
    \centering
    \includegraphics[height=0.12\textheight]{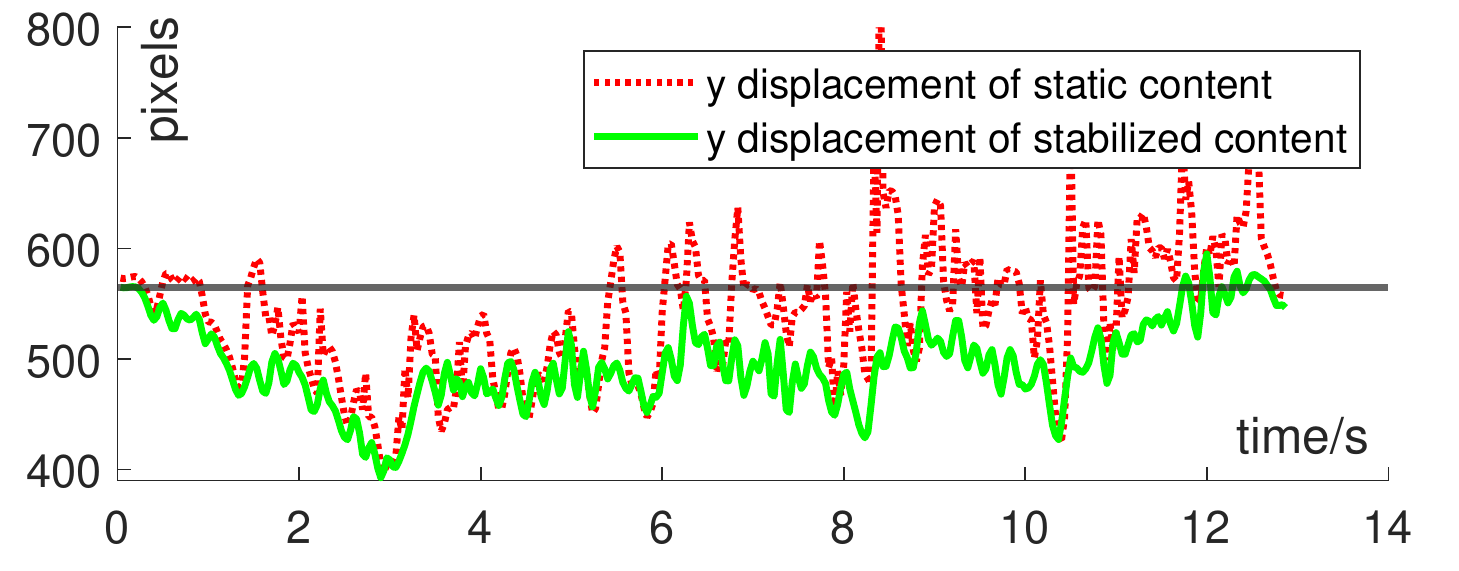}
    \vspace{-1ex}
    \caption{Y-axis (vertical) movement measured by a camera located 20 cm from our \name prototype while the latter was hovering.}
    \label{fig:movement}
\end{minipage}
\hfill
\begin{minipage}{0.45\textwidth}
    \centering
    \includegraphics[height=0.12\textheight]{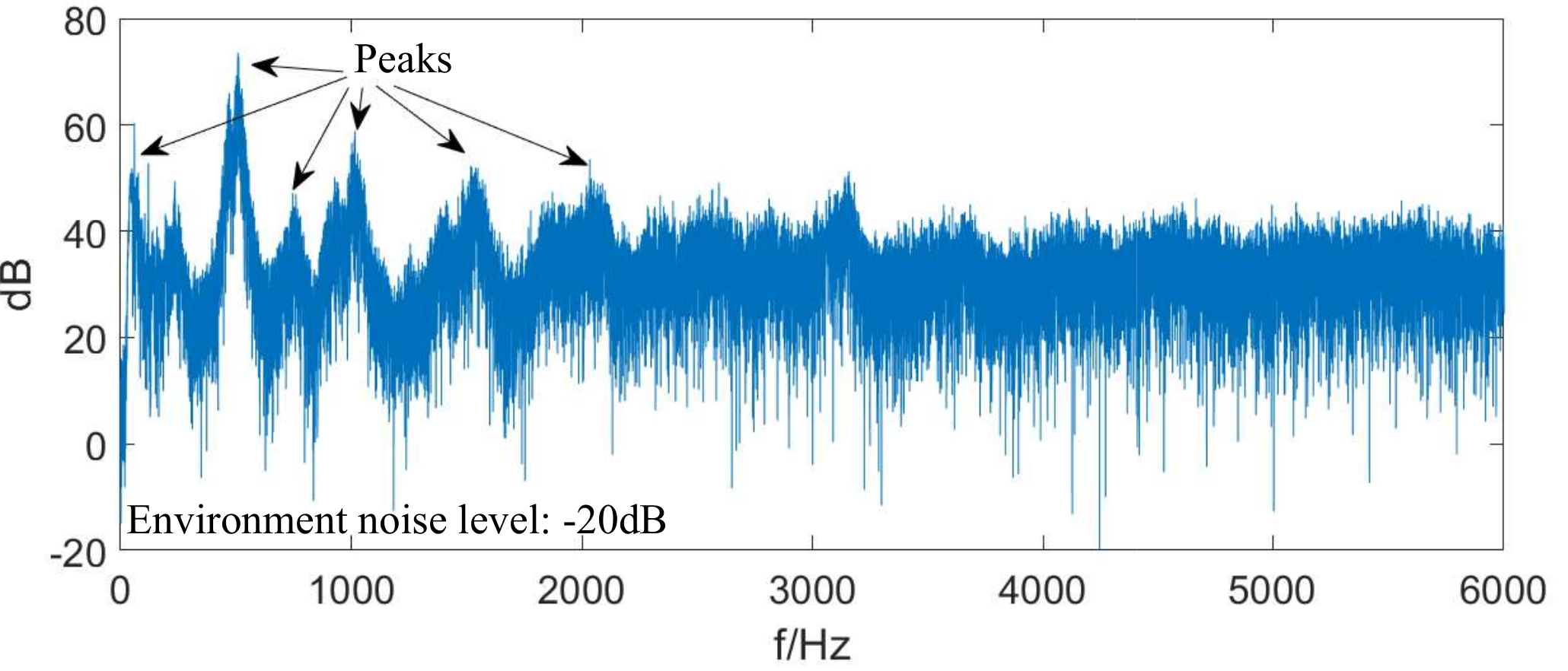}
    \vspace{-1ex}
    \caption{Noise spectrum of our \name prototype. It has multiple strong tonal components (peaks) and a relatively wide band.}
    \label{fig:noise_spec}
    \end{minipage}
    \vspace{-1ex}
\end{figure*}

We first describe several compelling applications of \name, a smartphone that flies. Because \name carries cameras and microphones, it can fill roles intended for existing cinematographic drones. We ignore these use cases and instead focus on those uniquely enabled by \name's display.
We envision that \name will share much of the hardware and software of smartphones. Our prototype uses a Google Pixel 4, allowing it to tap into the mature ecosystem of Android, which facilitates app development and enables ``retrofitting'' existing smartphone apps for \name. This flexibility contrasts with other drone platforms that develop their own ecosystems, e.g., DJI and Skydio. 

%Hands-free computing and communication: smartphone usage without hands

\textit{Hands-free ``Smartphone''}:~~
\name's hands-free experience opens up a host of new multitasking opportunities for mobile users in situations when holding a device in hand is either inconvenient or impossible. Users could make video calls or consume media while cooking, walking around the house, or jogging. 

%Friendly human-drone interaction with Display
\textit{Personable Interaction}:~~
There is a growing literature about human-drone interaction with a focus on how users command a drone~\cite{cauchard2021hdi}, e.g., using gestures~\cite{cauchard2015drone,obaid2016gesture,peshkova2017natural} and touch controls~\cite{abtahi2017touch}. Few works, however, examine how a drone could communicate with users directly, without using an intermediary device like a handheld terminal. For example, Szafir, Mutlu and Fong~\cite{szafir2015communicating} studied how a drone could use an LED ring to communicate its flight intention.  

\name, with its display, provides a much richer platform for this type of direct communication. For example, \name can use smiley faces to comfort the user or indicate its intention, which is important because many users are uncomfortable with a drone flying around at such close proximity.
\name can also use arrows or other symbols to direct the user's attention.

\textit{First Response and Public Safety}:~~
%First responders
We also envision \name being used in the context of first response and public safety. Its display can present instructions to first responders as they navigate difficult scenarios, like fires and disasters. It could present interactive maps or other informative graphics to those trapped in hard-to-reach areas.
At the scenes of car accidents, police could send \name to facilitate a remote video call with the driver and inspect their license, or to review damages and interview witnesses on the scene. 

% \name could also help to control large, noisy crowds effectively. If some danger is presented to the crowd, \name could fly to different sections of the crowd and present different directions, which is difficult for other means such as loudspeakers.
% \lin{Pod's screen might be too small for this purpose? We really need "personal" situation where information needed to be delivered to one or a few people.}
% \guojun{In public place, POD could serve as the food order robot. If anyone wants to order some food in airport, he could engage with a POD. Then if a user has ordered the food and waits for it now, the drone could fly to the customer and tell him the food is done. If POD has more capacity, it can even deliver the food.}
% \lin{Why does it need to fly? Sounds like a robot on wheels would suffice. The keys of POD are: (1) it can display personalized information at a close distance; (2) it can fly in order to overcome obstacles that are difficult for robots on wheels or legs.}

% %Public information
% Another use for our flying smartphone platform is to better inform the public. POD could serve as an informational kiosk. Similar to a jumbotron in Time Square, New York City, it could display news, alerts, and other content while offering freedom of movement and using computer vision to intelligently relocate itself relative to crowds or vehicles.

% \lin{Instead of jumbotron, I think a more compelling case would be a police sending a POD to ``interview'' a suspect, e.g., at a traffic stop. The display is critical here to add a human face.}

% !TEX root = ./main.tex 

\section{Open Challenges}
\label{sec:challenge}

While drones for cinematography have been extensively studied and widely used,
drones as a personal device have not. Our experience with the \name prototype has revealed technical challenges unique to personal drones, especially those related to their interacting with the user at such a close proximity.

\subsection{Display Usability}
%% shake of the drone, quantify the shake

Drones are dynamic systems. Keeping them stationary in the air is a long-studied and difficult problem. In order to capture high-quality videos and photos, commercial cinematographic drones have gone a long way toward solving this problem. Unlike \name, they benefit from two factors: they usually hover far away from the target scene; and the impact of instability on the footage they take can be removed by post-processing.

\name, on the other hand, must present a stationary, readable display to its user, and at a much closer distance, with the screen content often being generated in real time. \autoref{fig:movement} (red, dotted line) presents how the screen of our prototype \name moves along the vertical axis during one of its flights, despite its best effort to hover, as captured by a camera.  Such motion poses significant usability challenges for the display, especially for reading small text. Using a drone-carried iPad, the authors of~\cite{schneegass2014midair} found, not surprisingly, that a moving display carried by the drone requires a larger text font for readability, although the study did not isolate the impact of the drone's inability to stay stationary in the air, as is important to \name. 

\subsection{Noise Suppression}

Drones are noisy, and drone noise suppression has been studied extensively~\cite{miljkovic2018methods}. The noise worsens disproportionately when the user is close to the drone. And \name intends to serve its user at the intimate smartphone distance, much more closely than cinematographic drones, which acerbates the noise problem. 
Our \name prototype, though quite small, can produce noise up to 75 dBA, if measured from two feet away.
Commercial cinematographic drones are about as noisy, or noisier, due to their larger size~\cite{droneNoise}. 
Noise this loud is known to be harmful of humans during long periods of exposure~\cite{NoiseCDC}.
Cinematographic drones usually operate far from users, and noise captured by their microphones can largely be removed through post-processing. In contrast, \name operates at the smartphone distance from the user and the noise must be suppressed in real-time.

Many have explored using passive methods to reduce drone noise: covering the propellers with sound-absorbing materials and metal surfaces~\cite{mohamud2018drone}; using ducted (or shielded) propellers \cite{Malgoezar2019}; using specially shaped propellers, e.g., DJI low-noise propellers~\cite{DJILP}; and using propellers with more blades. 
These methods have only limited success. For example, ducted propellers~\cite{Malgoezar2019} only reduce the noise by up to 7 dBA.

The synchrophaser system \cite{james2012noise} is an active drone noise cancellation method. It requires that all propellers have exactly the same rotational speed and a fixed phase difference. It thus only works for fixed-wing aircraft with symmetric sets of engines. However, small rotorcraft like drones do not meet the synchrophaser requirements\textemdash their rotors constantly change speed to keep the craft stable.
Noise cancellation techniques widely used in headphones can reduce noise at the ear. These techniques work well with low-frequency ($<2$ KHz) and tonal noise.
Unfortunately, as shown in \autoref{fig:noise_spec}, drone noise consists of tonal noise of both high and low frequencies, as well as random wide-band noise higher than $6$ KHz. While active noise cancellation (ANC) has been implemented to reduce extra noise captured drone microphones~\cite{Wang2020, chun2019drone},
no existing drones employ ANC to reduce the noise heard by its user, as is the goal of \name.

%%%% comparison between remote tracking and close-range tracking
\subsection{User Following}
\name needs to present its display to a potentially mobile user. We envision that it mimics a human assistant holding the smartphone for its owner. This natural following model goes beyond the following capability enjoyed by commercial drones or reported in the literature. 
First, \name has to not only \emph{follow} the user but also orient the display properly. 
Second, while it is tempting (and feasible) to position \name at a constant distance with a constant orientation relative to the user's head or shoulders while the user moves around, this can be annoying. Imagine a human assistant holding the smartphone for the assisted. The assistant might decide to remain stationary while the assisted turns away from the display temporarily. One could imagine there exist ``eager'' vs. ``lazy'' assistants, with a lazy assistant preferring to stay stationary and following the assisted only ``reluctantly''.  \name must allow its user to adjust how eagerly it should behave.

% Nowadays, some commercial drones have the ability to track the user. There are two kinds of tracking according to the use case, long and short range tracking. Long range tracking refers to the situation where the drone is far away from the user, for example, cinematography, landscape photography, and extreme sport shooting. In these use cases, the drone usually hover high and track user from distance, so the delayed and inaccurate positioning won't be perceivable by user. However, short range human tracking have to realize more than keeping the user inside the picture. Some selfie drone now can do human recognition and tracking, but in a relative 
% static situation with human facing towards the drone directly. As the distance between the drone and user decreases, the inaccurate positioning will be more abrupt and easily noticed by the user.
% \name here wants to realize not only static user tracking, but also timely and fast mobile user following. 

\subsection{Programming Interface}
\name shares much of the hardware and software of smartphones and can benefit from the mature ecosystems of smartphones in app development. However, the drone's movement poses fresh programming challenges to even skilled developers. We consider two different programming support features. First, a \name application that can retrofit existing mobile apps. This retrofitting application would provide a natural interface for user input. It would not only convert speech or gestures into touch commands but also control \name's movement. For example, with this app, a \name user can use Android's built-in Camera app to take a photo after directing \name to the right position. 

Another, more interesting, support feature would be a library that developers can use to command \name and create novel apps catered to \name's use cases, like those outlined in \S\ref{sec:use}. Both the capabilities of the drone and the latency of the phone-drone communication constrain this library interface.
For example, a programming interface that directly sets the drone's motor thrust is unsafe because the drone firmware itself stabilizes \name and must have complete control over the motors. And the (non-deterministic) delay of USB communication between the phone and the drone makes it challenging for the phone itself to implement the stabilizer. 
Similarly, a programming interface that sets the 3D position of \name is infeasible because the drone itself is unable to accurately localize itself.
\section{Design and Implementation}\label{sec:system}
We next present our early prototype of \name, which is based on a custom-made nano-drone and a Google Pixel 4 smartphone.
A video demo of our prototype interacting with a remote-controlled, life-size mannequin is available at \cite{ourFollowingVideo}.

\subsection{Mechanical \& Electrical}

\textit{Drone}:~~
    \name's drone consists of a controller board, motor components, and a 3D-printed frame. We base the controller board on the Bitcraze Crazyflie, an open-source nano-drone platform~\cite{crazyflie}. 
    We upgraded Bitcraze's board to have more I/O ports. Unlike the Crazyflie, which uses brushed motors, \name uses 1104-4500KV brushless motors and an iFlight 12A electronic speed controller (ESC). We modify the Crazyflie firmware's motor control code to accommodate the ESC. We add a height sensor and an optical flow sensor to the drone in order to achieve autonomous hover stabilization (See \S\ref{sec:hover}).
    The drone is powered by its own dedicated battery and has a flight time of four minutes when carrying the phone.
    
\textit{Smartphone}:~~
        Our prototype uses a Google Pixel 4. We remove all unnecessary parts to pare down the mass as much as possible. Only the motherboard, USB-C connector, display, and battery are kept for our experiment, totalling a mass of 63 grams.
        The Pixel, which can act as a USB host device, sends control packets to and receives data from the drone via one USB-C to Micro-B cable.

% \begin{figure}[t]
%     \centering
%     \setlength{\belowcaptionskip}{-10pt}
%     \includegraphics[width=0.8\linewidth]{image/sys-diag.pdf}
%     \caption{System diagram of \name}
%     \label{fig:sys_diag}
% \end{figure}

\subsection{Autonomous Hover}
\label{sec:hover}
%% drone stabilization
Smaller drones, including the Crazyflie, do not usually come with stable autonomous hovering out of the box; a human needs to manually direct the hover via remote control. But several methods are available to enable autonomous hover. Most cinematographic drones localize themselves with GPS, which only has meter-level precision
and only works outdoors. Some drones use a motion capture system consisting of multiple cameras to precisely calculate drone location within the coverage area, e.g.~\cite{FMA, kushleyev2013towards}. Such motion capture systems are not only expensive but also have limited coverage area, not suitable for the use cases of \name. 

Our \name prototype uses a much simpler, more mobile, solution: an optical flow sensor and phone camera. Optical flow is the pattern of apparent motion of objects due to the relative motion difference between the observer and a scene. By tracking the optical flow of the ground, a drone can know its relative horizontal displacement and make timely adjustments. We use the phone camera as a secondary observer to make the drone aware of its global environment. Unlike the optical flow sensor, the camera can help the drone stay within the user's sight. 

\begin{figure}[t]
%\centering
%\begin{minipage}{0.45\textwidth}
    \centering
    \includegraphics[width=0.8\linewidth]{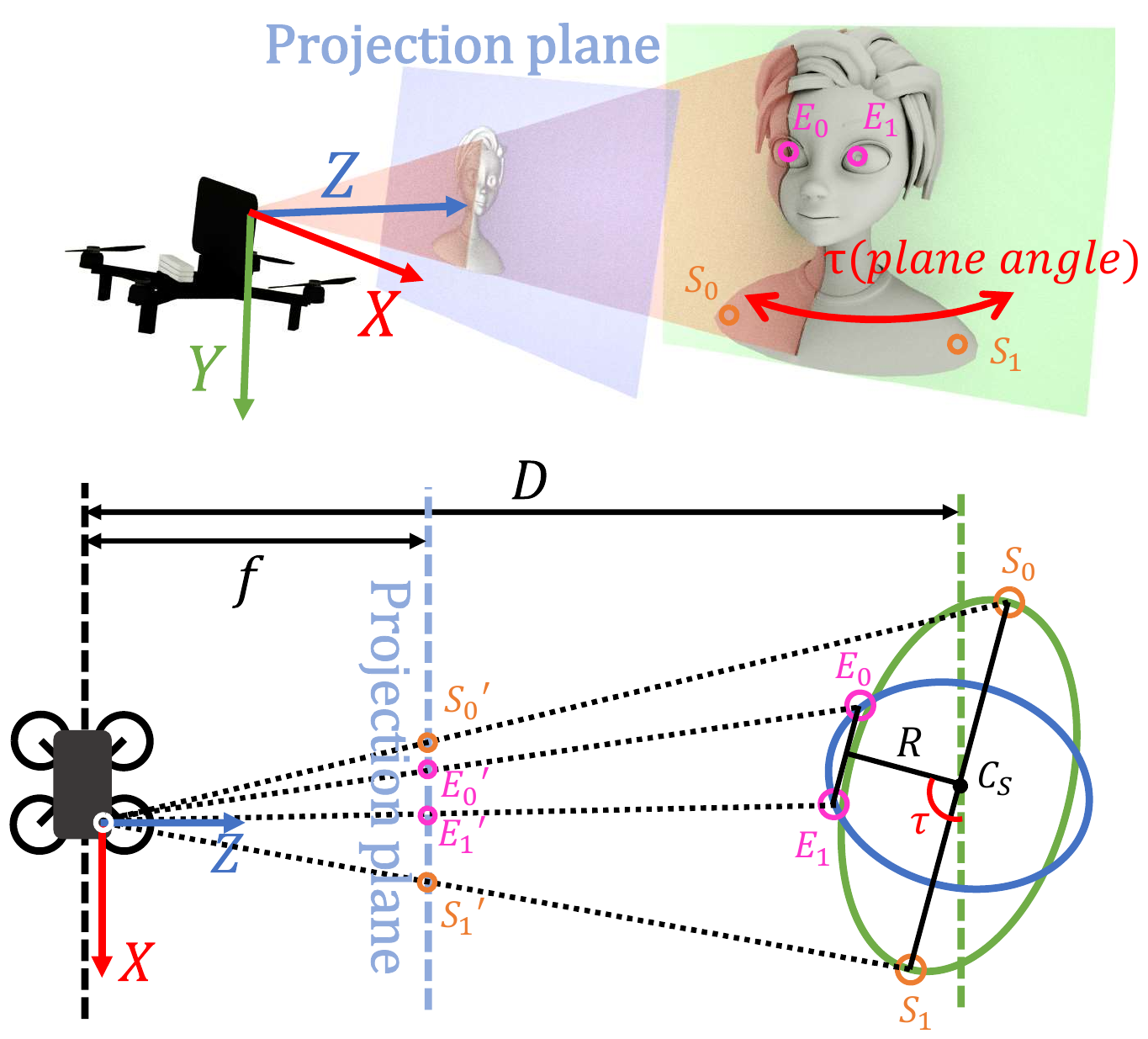}
    \caption{Spatial relationships between \name and its user: side view (top) and top view (bottom). In the side view, the red plane is part of the camera's X plane, and the green plane is the plane that passes through the shoulders, i.e., $S_0$ and $S_1$, and is perpendicular to the ground. $\tau$ represents the angle between the camera X plane (red, parallel to the plane defined by the Y and Z axes) and the user's shoulder plane (green). The projection plane is parallel to the X-Y plane and lies at a distance of $f$ from the camera, where $f$ is the camera's focal length. }
    \label{fig:spatial}
%\end{minipage}
%\hfill
% \begin{minipage}{0.45\textwidth}
%     \centering
%     \setlength{\belowcaptionskip}{-10pt}
%     \includegraphics[width=1\linewidth]{image/torso-birdseye-cropped.pdf}
%     \vspace{-2ex}
%     \caption{Top view of the user's head (blue) and shoulders (green). Y-axis is perpendicular to the plane of the paper. $p$ and $q$ change as the user changes their orientation $\tau$.}
%     \label{fig:topview}
%     \end{minipage}
\end{figure}
\subsection{Vision}
\label{sec:vision}

\name uses computer vision to locate and track its user, leveraging the phone's front-facing camera and computational power, namely through PoseNet.

% \begin{figure}
%     \centering
%     \setlength{\belowcaptionskip}{-10pt}
%     \includegraphics[scale=0.15]{image/posenet_demo.png}
%     \caption{A sample frame, with human feature keypoints found by Posenet marked by red dots, and our estimation of angle $\tau$ indicated by the axes diagram.}
%     \label{fig:posenet}
% \end{figure}

\textit{PoseNet}~\cite{posenet} is based on PersonLab \cite{papandreou2018personlab}, a fully convolutional neural network implemented in TensorFlow. 
%The network is trained in a supervised manner on the COCO dataset, which annotates 17 ground truth keypoints: 12 for the body and five for the face. 
%The network, built from a modified ResNet-based CNN backbone, uses a bottom-up approach, meaning that instead of starting by finding humans in the frame using a bounding box object detector, it starts by finding identity-free semantic features in the frame (i.e., labeled semantic image segments) and proceeds to group them into person instances. PoseNet is trained in a fully supervised manner on the COCO dataset, which annotates 17 ground truth keypoints: 12 for the body and other 5 for the face. 
%It takes an input image of any size and returns the 17 human keypoints in the form of one 3D heatmap tensor and one 3D offset vector tensor. The heatmap tensor (image dimension x image dimension x 17) contains the probabilities that each pixel contains a part of the given keypoint type. The offset tensor gives the estimated x and y offsets from each pixel to arrive at the center of the joint or feature keypoint.
Because PoseNet is computationally heavy, we reduce its use with the Lucas-Kanade (LK) optical flow algorithm. The algorithm matches the human feature points of a frame with the same corner feature points from the previous frame and  estimates  the  geometric  displacement  between  these  two sets of points. Because the LK algorithm is much more efficient than PoseNet, we invoke PoseNet to retrieve accurate feature points only every fourth frame and use the LK algorithm in between to track these points. 

% We present a simple method for estimating the 3D position of the human relative to the camera. We assume that the Pixel's camera satisfies the pinhole camera model, which defines the relationship between a 3D point $[X, Y, Z] \in \mathbb{R}^3$ in the camera coordinate system and a 2D pixel $[x, y]^T$ in the image frame, yielding

% \begin{equation}
% \begin{bmatrix}
% x\\
% y\\
% 1\\
% \end{bmatrix} = 
% \begin{bmatrix}
% f_x & 0 & x_0 & 0\\
% 0 & f_y & y_0 & 0\\
% 1 & 0 & 1 & 0\\
% \end{bmatrix}
% \begin{bmatrix}
% X\\
% Y\\
% Z\\
% 1\\
% \end{bmatrix}
% \end{equation}

% where $f_x$ and $f_y$ are the focal lengths of the camera in the x and y directions, and $[x_0, y_0]^T$ is the optical center of the camera. We assume that $f_x$ and $f_y$ are both equal to some focal length $f$, and that the optical center is equal to the center point of the captured images.
% Since computing the position and yaw angle of the human only requires a few real known measurements, we do not need to calibrate the camera. We can discard the conception of a world coordinate frame, assuming that the camera and world share the same coordinate system.

\vspace{1ex}\textit{Human Position and Orientation Estimation}:~~
\name must estimate the user's distance and orientation in its own coordinate space (or camera coordinate space) \emph{without} using a depth camera.
\autoref{fig:spatial} depicts the spatial relationship between \name and its user. The distance and orientation to be estimated are marked as $D$ and $\tau$ respectively.
We assume the following parameters are known, either from calibration or being directly supplied by the user: (\textit{i}) the focal length $f$ of the camera; (\textit{ii}) the distance, $L_e$, between the user eyes, $E_0$ and $E_1$ in \autoref{fig:spatial}; (\textit{iii}) the distance, $L_s$, between the user's shoulders, $S_0$ and $S_1$ in \autoref{fig:spatial}; and (\textit{iv}) the distance between the two parallel planes perpendicular to the ground defined by the shoulders and the eyes, respectively, marked $R$ in \autoref{fig:spatial}.

\name first measures the lengths of $S_0'E_0'$ and $S_1'E_1'$, denoted by $p$ and $q$, respectively, on the projection plane, by simply counting pixels in the camera view. 
When the user changes their orientation, $p$ and $q$ change accordingly. 
\name derives $\tau$ from $p/q$ using the following relationship:

\[ \frac{p}{q}= 
\frac{\frac{(L_{s}-L_{e})}{2}\sin\tau+R\cos\tau}{\frac{(L_{s}-L_{e})}{2}\sin\tau-R\cos\tau} \]

\name computes $D$ using the pinhole camera model as 
\[D = f \cdot L_e\cdot sin(\tau)/L_e'\] 

where $L_e'$ is the distance between the user eyes on the projection plane, i.e., $E_0'E_1'$. 

\vspace{1ex}\textit{Feedback control}:~~
To accommodate many useful tasks, \name must remain at a predetermined distance and orientation (yaw) relative to the user. To maintain these requirements in real time, we employ a set of three PID controllers that are synchronized with PoseNet, running at around 50 Hz. The controllers correspond to \name's yaw, distance from the human, and position along the x-axis.

\section{Early Results}\label{sec:evaluation}
Based on our prototype described above, we next present some early results in addressing the challenges discussed in \S\ref{sec:challenge}, with mixed success.

\textit{Evaluation setup}:~~In order to test the prototype, we mount a human bust mannequin on a robot tank chassis driven by two DC motors, as shown in \autoref{fig:podamable}. The mannequin has realistic features and is the actual size of a human torso\textemdash approximately 37 cm wide and 42 cm high. We use an Arduino and an ESP-01 Wi-Fi module to create a Wi-Fi remote control for the mannequin. The mannequin is recognized normally by PoseNet and contains all points of interest needed by our implementation (shoulders and eyes). Controlling the mannequin allows us to safely test our human following implementation, including much of our state machine.

\subsection{Display Usability}
\label{sec:stable}
%% vertical shake vs horizontal shake
In order to improve the usability of \name's display, we adjust the display content in real time to counter the movement of the drone. A naive implementation would use a sensor (camera or IMU) to measure the device's movement and inform the graphical rendering system to adjust content accordingly. However, the latency between sensing the movement and showing the adjusted content on the display would render this implementation inefficient and ineffective. Therefore, we experiment with several predictive stabilization methods originally intended for handheld devices, like NoShake~\cite{rahmati2009percom}. NoShake was intended to stabilize content when a user reads a handheld display while walking. Walking causes hand and, as a result, display movement relative to the eyes. NoShake combines a spring-mass model of the display content and readings from the phone's accelerometer to move the display content opposite to the display movement so that it remains stable relative to the user's eyes.

Our experimental measurements result in \autoref{fig:movement} shows that our current implementation of NoShake noticeably reduce the displacement of screen content when compared side-by-side with identical unstabilized content. 
Because movement of a hovering drone is different from that of a mobile user hand, i.e., less periodic and more frequent, we had to manually tune the spring factor and dampener friction constant to be effective as shown in \autoref{fig:movement}. We expect a data-driven, machine learning approach would be more effective. 
We also notice that when NoShake moves the screen content, it occasionally introduces the ghosting effect \cite{ghosting}, partly due to the very high refreshing rate of Pixel 4's screen ($90$ $Hz$).

\subsection{Noise Suppression}
Because prior work~\cite{mohamud2018drone,Malgoezar2019} has highlighted the limits of passive noise reduction, we focus on active methods.
Active noise reduction leverages destructive interference: it generates a sound wave with the same amplitude as the target noise but with inverted phase, such that the two cancel each other out at the receiver. In \name's case, the user's ears represent the receiver.
\begin{figure}[t]
    \centering
    \setlength{\belowcaptionskip}{-10pt}
    \includegraphics[width=0.8\linewidth]{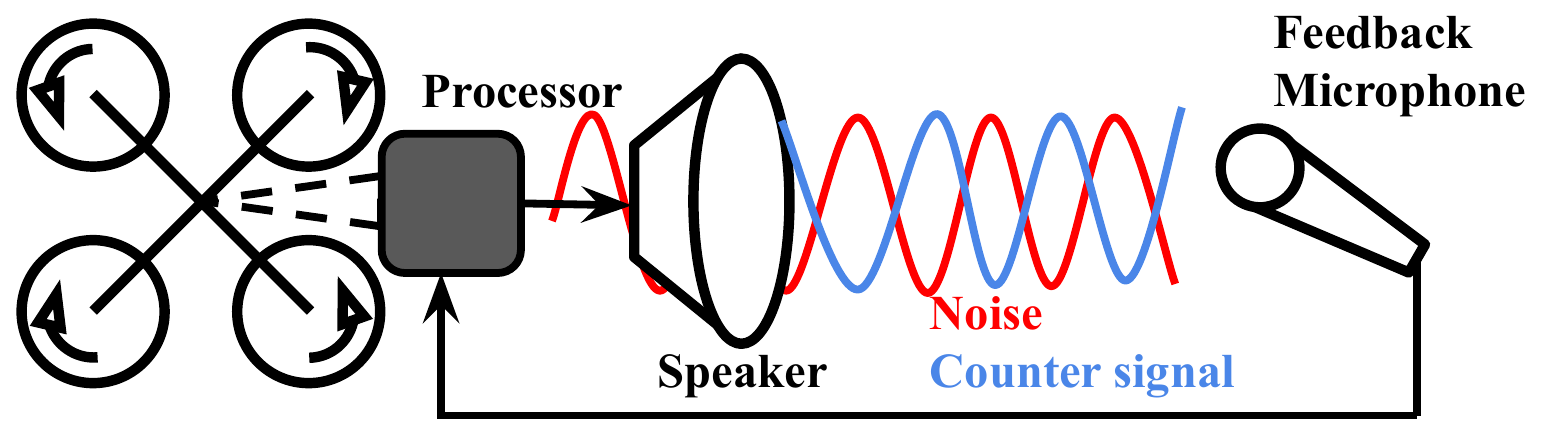}
    \caption{Our active noise reduction system consists of a speaker, a feedback microphone located near a user ear, and a processor on the drone. }
    \label{fig:noise_diag}
\end{figure}

We experiment with the feedback-based active noise cancellation setup show in \autoref{fig:noise_diag}, which requires a feedback microphone to be placed next to the user's ear (two feet away from \name). 
We focus on cancelling the prominent tonal components of the drone noise.
First, a micro-controller (STM32F4, similar to the one used in the drone) that is connected to the feedback microphone analyzes the noise spectrum to determine the tonal noise with the highest intensity.
Next, the speaker carried by the drone generates a sound wave with an adjustable phase and the same frequency as the high-intensity noise. The microphone provides feedback to the speaker, which can then adjust the wave's phase to perfectly cancel the noise.  

A decibel meter next to the feedback microphone shows that the above active cancellation technique reduces the noise from 73 dBA to 70 dBA. While small, this reduction is promising since we only used a single speaker and only cancelled the strongest tonal component. 
We plan to investigate the use of multiple speakers. Most noise cancellation solutions~\cite{zhang2018spatial} use multiple speakers and cancel more frequency components. 

\subsection{User Following}
\label{sec:follow}

\begin{figure}[t]
    \centering
    \includegraphics[width=1\linewidth]{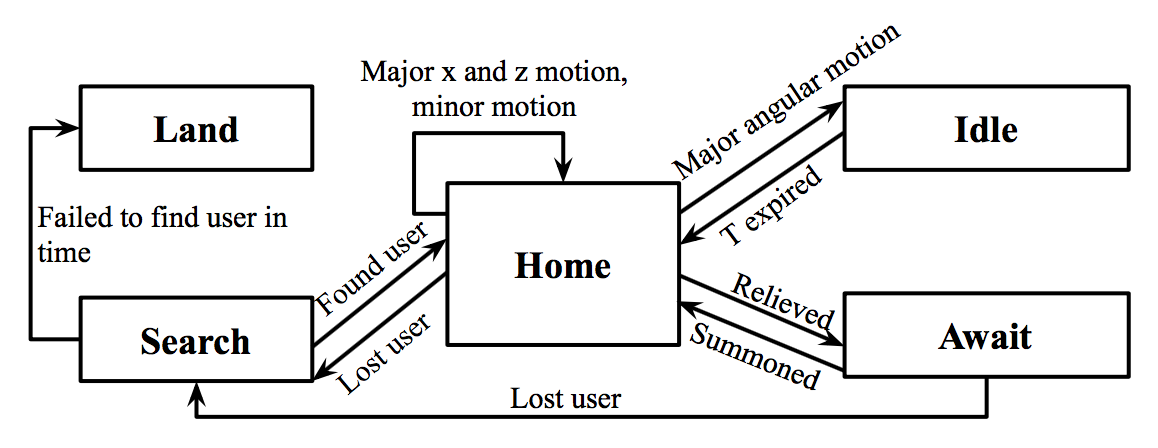}
    \caption{The behavioral model of \name. State transition is based on the perceived state of the user. In \dstate{Home}, \name actively follows the user and presents its display at a pre-determined distance and angle relative to their shoulders. In \dstate{Idle}, \name holds still. In \dstate{Await}, \name yaws to keep the user in the frame.}
    \label{fig:vision_diag}
\end{figure}

We experiment with configuring \name to mimic a human assistant holding a smartphone for the assisted. \autoref{fig:vision_diag} represents \name's behavioral model as a state machine. 
\name changes its state based on its understanding of the user's state.
\autoref{tab:humstaterec} summarizes what user states \name recognizes and how. The behavioral model incorporates a tunable parameter, $T$, which can range from a fraction of a second to tens of seconds, depending on the application. A larger $T$ leads to a ``lazier'' \name that follows its user more reluctantly.

In \dstate{Home}, \name actively maintains a predetermined distance and orientation with respect to the user's shoulders, using the feedback controller described in \S\ref{sec:vision} to follow the user when they make major movements as outlined in \autoref{tab:humstaterec}. \name ignores any minor movements. In \dstate{Idle}, \name does not move at all. \name enters \dstate{Idle} after the user rotates past the predetermined $\tau$ threshold. It returns to \dstate{Home} after $T$ elapses. 

The user can use the relieving and summoning gestures to indicate that it would like \name to immediately enter \dstate{Await} or re-enter \dstate{Home}, respectively. In \dstate{Await}, \name only yaws to keep the user in its camera view. That is, it only keeps the feedback controller for yaw on.

% \begin{figure}
%     \centering
%     \setlength{\belowcaptionskip}{-10pt}
%     \includegraphics[scale=0.12]{image/drone-model-cropped.png}
%     \caption{A behavioral model of the drone}
%     \label{fig:vision_diag}
% \end{figure}

\begin{table}[!t]
\scriptsize
\centering
%\resizebox{\textwidth}{!}{
\begin{tabular}{|p{0.08\textwidth}|p{0.31\textwidth}|}\hline 
\textbf{User State} & \textbf{Recognition Method} \\\hline 
 %\ustate{Engaged} & $\displaystyle \tau $ hovers near 0 for time $\displaystyle T_{g}$, head and shoulders roughly aligned \\\hline 
 %\ustate{Disengaged} & $\displaystyle |\tau |$ hovers above threshold $\displaystyle \tau _{g}$ for time $\displaystyle T_{g}$, head and shoulders roughly aligned \\\hline 
 \ustate{Summoning} & Raise right wrist above eye line \\\hline \ustate{Relieving} & Raise left wrist above eye line \\\hline
 %\ustate{Mobile} & At least one of X, Z, or angular velocities has positive magnitude \\\hline 
 \ustate{Major\_motion} & At least one of X, Z, or orientation exceeds acceptable ranges  \\\hline 
 \ustate{Minor\_motion} & No change in position or orientation that exceeds acceptable ranges, head begins to move while shoulders still or vice versa \\
 \hline
 \ustate{Lost} & No user in view\\\hline
\end{tabular}
%}
\caption{\label{tab:humstaterec}Methods for recognizing the current human state.}
\end{table}

% \lin{What's the difference between Mobile and Major\_motion for user states?}\noah{Mobile is a state the user moves into whenever they begin moving. Major\_motion is essentially a state the user moves into when they've been moving for some time. Unfortunately the table was supposed to go with the human model diagram I had pasted and described earlier, which had a cascading nature see link %https://www.mathcha.io/editor/KlX95U3ZioDU0vi5J33KeFMNMo1muwPmZ6vtO32lyz. 
% For example, major motion will happen WITHIN the engaged state, etc.}

A short video showing our prototype eagerly following our test mannequin is available here~\cite{ourFollowingVideo}. 

\subsection{Programming interface}
Our prototype's phone uses the Android USB library to control the drone via a connecting cable.
The drone used in our prototype can only determine its height (z-axis) accurately. It can estimate changes in the x-y plane with the optical flow sensor with less accuracy.

We currently support a programming interface allowing an Android app to directly command \name's absolute and relative movement along the z-axis (height) and its relative movement in the x-y plane. These functions are complete in that an \name app developer can use them to command \name to any position relative to its user, especially with feedback from \name's camera-based human position and orientation estimation (\S\ref{sec:vision}).
Implementing this USB programming interface requires a new periodic task in the drone firmware. The task runs at 100Hz, which is one-tenth of the main controller loop frequency. 
We also plan to expand our interface by converting our human position and orientation estimation technique (\S\ref{sec:vision}) into an Android library. 

The programming interface is blocking, or synchronous. That is, once the phone app invokes an API, it will block until the drone completes the requested movement and returns a message. We choose this synchronous design because we believe the app should move the drone one command a time, sequentially, for safety. Asynchronous motion APIs could lead to unpredictable and dangerous behavior.
\section{Related Works}\label{sec:related}
% \name intersects with many works which explore the potential of the drone as a new approach to Human-Robot Interaction (HRI).

% \subsection{Drone Localization}
% Most outdoor drone applications \cite{Mueller2015, ohta2017sky} require GPS or real-time kinematic (RTK) GPS to realize fast and accurate positioning and navigation. However, they are robust only for outdoors, and the bulky module size makes them an improper choice for personal drone applications.

% \cite{Mao2017indoor} introduces an acoustic solution for indoor drone tracking application. It requires an exposed mobile phone to receive acoustic signal from the drone, and it has a very limited working range due to the quick attenuation of acoustic signal.
% It is much more common to adopt the motion capture system \cite{OptiTrack} for indoor drone localization. Though being high-precision and fast, such a system consists of multiple costly wide-range infrared cameras and only works in a certain areas that is covered by all of them.

\textit{Drone as User Interface}:~~
Many have explored harnessing drones to design new mobile user interfaces like flying 3D displays~\cite{rubens2015bitdrone}, haptic interfaces~\cite{Yamaguchi2016haptic}, and projectors~\cite{Darbar2019, Toyohara2017, Zhang2019,Scheible2013, Brock2018flymap}.
DisplayDrone~\cite{rubens2015bitdrone} carries a flexible touch display with remote tele-presence functionality. ISphere \cite{Yamada2017isphere} is a flying, omni-directional, spherical LED display. The authors of~\cite{schneegass2014midair} studied text readability on a tablet computer carried by a drone. 
These works were partially geared towards presenting information and guidance to mobile users. Their focus, however, was not on system design and implementation. Their prototypes offer no communication between the display content and the drone. In contrast, \name tightly integrates the drone and the phone and depends heavily on collaboration between them.

\textit{Human Following by Robots}:~~
There is a growing literature that studies how mobile robots can track and follow human users.
%~\cite{hu2013following,gupta2016novel,liu2018novel,honig2018socially,lee2018robust,islam2019person,boschi2020cost}. 
While \name builds on many ideas presented by past works, it uniquely features a rich user interface with both input and output flow. It thus faces a new set of challenges. Existing robotic followers, including drones~\cite{Mao2017indoor,boschi2020following}, usually aim at staying within a certain distance of their human users. \name must additionally orient the display properly and adjust screen content in real time as described in \S\ref{sec:stable}.
%\lin{discuss literature on non-drone robots presenting a display and following human users.}

%%% Related to vision-based following
%Following using human detection based on YOLO~\cite{boschi2020following}.

%%% Related to homing
% Related to the ``homing'' behavior of \name, Monajjemi, Mohaimenianpour, and Vaughan~\cite{monajjemi2016homing} studied how a distant drone could approach a user after detecting a specific gesture.

The Georgia Tech Miniature Autonomous Blimp (GT-MAB)~\cite{yao2017blimp} can follow and track a human user using a camera and multiple thrusters.
Unlike drones, blimps do not rely on motorized propellers for lift, so they are quieter, more stable and much easier to maneuver.
While GT-MAB and \name accept similar user input and both use vision-based sensing, GT-MAB offers no output or display. The primary challenge facing a blimp-based \name would be its size: to carry the same reduced smartphone system (63 gram) as used in our prototype, GT-MAB must gain a volume of $0.6$ m$^3$.

\bibliographystyle{acm}
{\scriptsize\bibliography{bib/abr-short, bib/zhong, bib/drone}}

\end{document}